%% file: elsarticle-template-num.tex
\documentclass[preprint,12pt]{elsarticle}

\usepackage{amssymb}

\usepackage{hyperref}       
\usepackage{url}            
\usepackage{booktabs}       
\usepackage{amsfonts}       
\usepackage{nicefrac}       
\usepackage{microtype}      

\usepackage{amssymb}
\usepackage{amsmath}
\usepackage{subcaption}
\usepackage{graphicx}
\usepackage{float}

\usepackage{fancyvrb}
\VerbatimFootnotes

\usepackage{placeins}
\usepackage{graphicx,psfrag,color} 
  \graphicspath{ {./images/} }
\newcommand{\sect}[1]{Section~\ref{#1}}
\newcommand{\fig}[1]{Fig.~\ref{#1}}

\newcommand{\equ}[1]{(\ref{#1})}

\newcommand{\K}{\mathcal{K}}
\numberwithin{equation}{section}

\newtheorem{remark}{Remark}[section]

\newcommand{\assign}{:=}
\newcommand{\tmem}[1]{{\em #1\/}}
\newcommand{\tmop}[1]{\ensuremath{\operatorname{#1}}}

\definecolor{myred}{RGB}{160,0,0}
\definecolor{mygreen}{RGB}{0,160,0}
\definecolor{myblue}{RGB}{0,0,160}
\hypersetup{
                colorlinks = true,
                linkcolor = {myred},
                citecolor = {mygreen},
                urlcolor  = {myblue},
}

\journal{Wave Motion}

\begin{document}

\begin{frontmatter}



\title{Diffraction by a set of collinear cracks on a square lattice: an iterative Wiener-Hopf method}


\author[inst1]{Elena Medvedeva \corref{cor1}}
\ead{elena.medvedeva@manchester.ac.uk}

\affiliation[inst1]{organization={Department of Mathematics of The University of Manchester},
            addressline={Oxford Road}, 
            city={Manchester},
            postcode={M13 9PL}, 
            country={United Kingdom}}

\author[inst1]{Raphael Assier}
\ead{raphael.assier@manchester.ac.uk}

\author[inst1]{Anastasia Kisil}
\ead{anastasia.kisil@manchester.ac.uk}

\cortext[cor1]{Corresponding author}


\begin{abstract}
The diffraction of a time-harmonic plane wave on collinear finite defects in a square lattice is studied. This problem is reduced to a matrix Wiener-Hopf equation. This work adapts the recently developed iterative Wiener-Hopf method to this situation. The method was motivated by wave scattering in continuous media but it is shown here that it can also be employed in a discrete lattice setting. The numerical results are validated against a different method using discrete Green’s functions. Unlike the latter approach, the complexity of the present algorithm is shown to be virtually independent of the length of the cracks.
\end{abstract}


\begin{highlights}
\item The numerical complexity is virtually independent of the cracks’ length
\item The iterative method applies for a mixture of finite and semi-infinite cracks
\item The iterative Wiener-Hopf method allows to avoid the matrix factorisation
\item The iterative method converges with a very modest number of iterations
\item The numerical complexity of an iteration does not depend on the iteration number

\end{highlights}

\begin{keyword}
discrete Wiener–Hopf equations \sep iterative methods \sep square lattice \sep finite crack \sep acoustic waves scattering
\end{keyword}

\end{frontmatter}


\input{1_introduction}

\input{2_problem}

\input{3_iterative}
\input{4_results}
\input{5_conclusions}

\appendix

\input{appendB}
\input{appendA}

 \bibliographystyle{elsarticle-num} 
 \bibliography{mendeley}





\end{document}

%% file: 1_introduction.tex
\section{Introduction}

\begin{figure}[ht]
    \centering
\begin{subfigure}{.3\textwidth}
    \centering
    \caption{Problem with a finite crack of a length $L$ in a lattice. The links between the red dots are broken (see \sect{sect:single finite crack}).}
    \includegraphics[width=\textwidth]{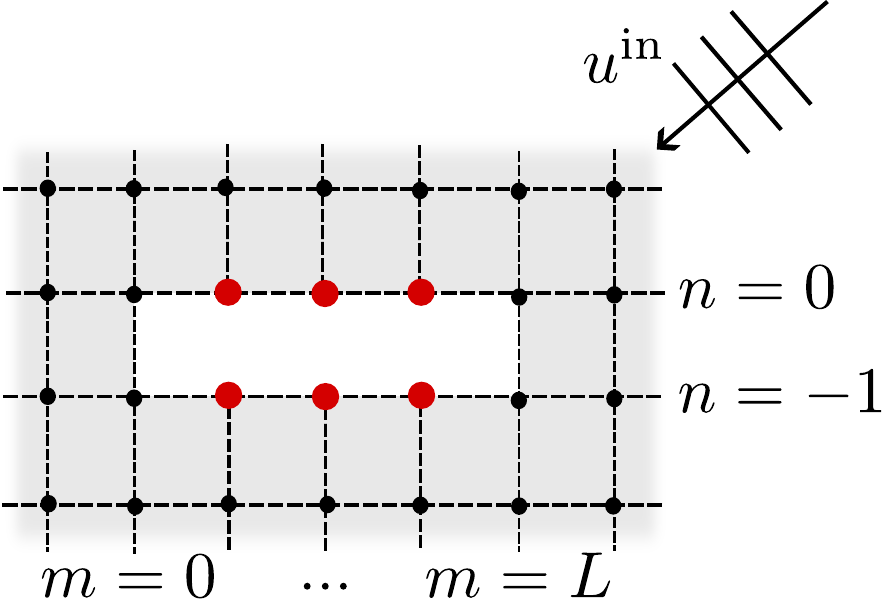} 
    \label{fig:crack_scheme}
\end{subfigure} 
\hspace{.15\textwidth}
\begin{subfigure}{.3\textwidth}
    \centering
    \vspace{.75cm}
    \caption{Problem with a finite barrier of a length $(L-2)$ in a lattice. The red dots are fixed (see \ref{appx:wh_rigid}).}
    \includegraphics[width=\textwidth]{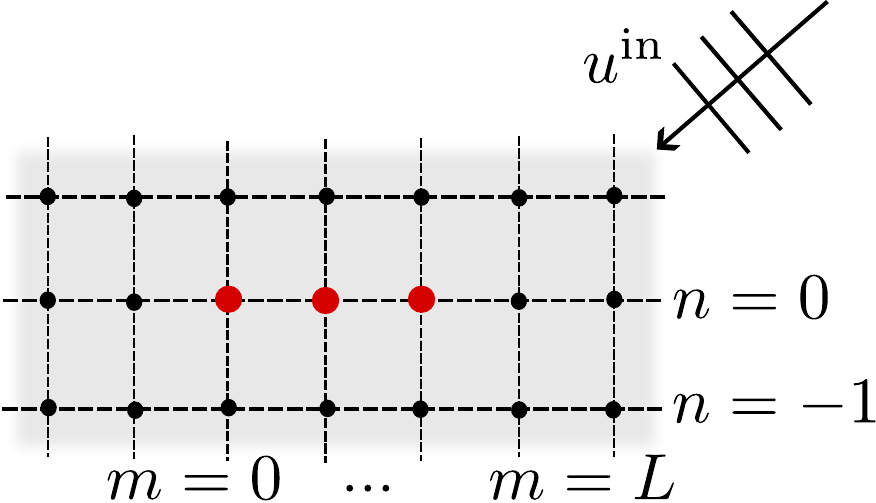} 
    \vspace{.25cm}
    \label{fig:rigid_scheme}
\end{subfigure} 
\begin{subfigure}{.5\textwidth}
    \centering
    \caption{Problem with collinear finite cracks in a lattice. The links between the red dots are broken (see \sect{sect:colliear_cracks}).}
    \includegraphics[width=\textwidth]{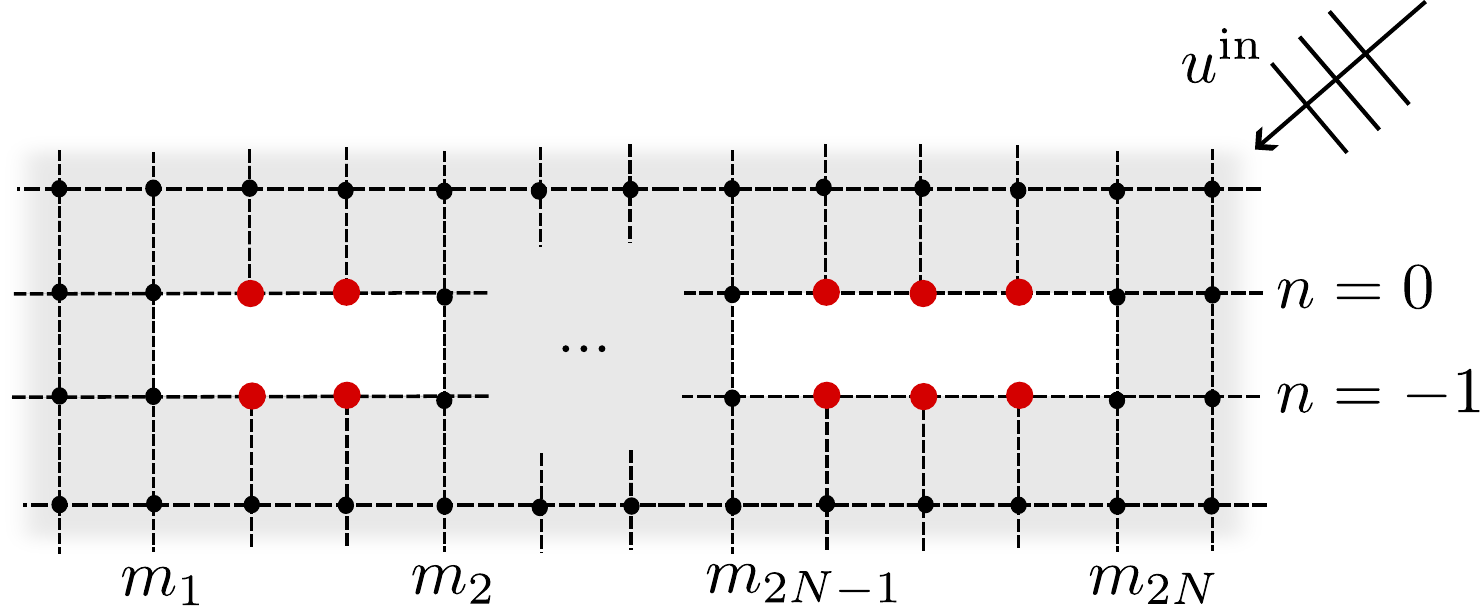} 
    \label{fig:cracks_scheme}
\end{subfigure} 
\caption{A scheme for the problem of a plane wave scattering by various square lattice effects.}
\label{fig:lattice deffects}
\end{figure}

A square lattice model is widely used in the literature to describe various processes in solid mechanics \cite{Slepyan2002ModelsMechanics, Mishuris2008DynamicsLattice} as well as waves propagating through elastic solid media \cite{Craster2023MechanicalMetamaterials,Vanel2016AsymptoticsFunctions}. In this work, we are going to consider the scattering of an anti-plane incident plane wave on finite defect structures in the lattice (\fig{fig:lattice deffects}), in particular, a finite crack (\fig{fig:crack_scheme}) and a set of collinear cracks (\fig{fig:cracks_scheme}).
The exact solution for such formulation was obtained by Sharma for a finite crack \cite{Sharma2015Near-tipCrack} and finite rigid constraint \cite{Sharma2015Near-tipConstraint}. Both cases are based on the usage of the discrete Green's function. This method has been further implemented in a series of works, such as \cite{Maurya2020WaveCracks}, where the diffraction of a plane wave on an infinite vertical array of staggered finite cracks is discussed. They use Floquet periodicity to reduce the problem to that of a specific wave-mode scattering by a single waveguide crack. 
A similar so-called boundary algebraic equations (BAE) method \cite{Martinsson2009BoundaryProblems,Gillman2010FastLattices} for the  boundary scattering problems with a finite scatterer on the square lattice is discussed in \cite{Poblet-Puig2015SuppressionEquations} and later in \cite{Kapanadze2018ExteriorLattice}. The continuous analogue of the method with the Green's function for the crack in a continuous medium is discussed by Wickham in \cite{Wickham1981TheExistence}, which can be considered a formulation of a well-known boundary element method (BEM).

However, the computational cost of this method is high, as for a crack of the length $L$
it requires inverting a matrix of dimension $(L-1)\times (L-1)$, each entry of which contains a lattice Green's function in the form of a double integral. Therefore, while for a small crack, we can calculate the solution using, for example, Green's function's far-field asymptotics \cite{Sharma2015Near-tipCrack,Martin2006DiscreteLattice,Shanin2024AStructures}, or evaluating the double integral numerically, it appears to be needlessly costly from a numerical perspective for a realistic crack of size $L\gg1$. The approach in the present paper is such that the computational cost is virtually independent of $L$.
In \cite{Sharma2015Near-tipCrack}, Sharma shows that in the case of a large crack, a semi-infinite approximation can provide a good approximation for the near-tip field. In this paper, we will improve on this by obtaining an arbitrary good approximation using iterations. Solutions for the semi-infinite defects can be obtained by the scalar Wiener-Hopf technique \cite{NobleB1958MethodsPDEs} for the discrete setting, which is a direct analogy of the Sommerfeld problem in a continuous medium (see \cite{Sharma2015DiffractionCrack,Sharma2015DiffractionConstraint}).

Unlike the semi-infinite defect, which leads to the scalar Wiener-Hopf equation, the case of a finite crack can be formulated as a matrix Wiener-Hopf equation. The kernel of this equation is a square matrix with growing factors on its diagonal. The factorisation of matrix functions is generally an open question, with only specified classes possessing constructive methods \cite{Rogosin2016ConstructiveMatrix-functions,Kisil2021TheMethods}. In some cases, such as two staggered semi-infinite plates, the matrix can be reduced to a system of scalar linear equations -- see works 
\cite{Abrahams1988OnFactorization,Abrahams1990TheMode,Abrahams1990AcousticPlates,Abrahams2006GeneralFactors} for continuous setting, \cite{Maurya2019ScatteringFactorization} for the discrete analogue, and \cite{Sharma2020ScatteringZone} for the case of a finite damage zone in the lattice. 


In \cite{Kisil2018AnFactors},\cite{Kisil2018AerodynamicExtensions}, an iterative Wiener-Hopf method has been suggested for a class of triangular matrix functions. This approach can be considered as a generalisation of the `pole removal technique' and Schwarzschild’s series \cite{Schwarzschild1901DieI.Schwarzschild}. The method is proven to be convergent and reaches sufficiently high accuracy with only a few iterations. In \cite{Priddin2020ApplyingFactors}, the method has been generalised for solving $n\times n$ matrix Wiener-Hopf equations, i.e. for the problem of scattering by a set of collinear `soft' or `hard' plates. In \cite{Bharth2022MathematicallyCylinder}, the iterative method has been adopted for the problem of deformation in an elastic half-space, caused by a cylindrical roller, in \cite{Nieves2022DynamicChains,Livasov2019NumericalZone}, a variation of the iterative method has been used for studying crack propagation. The present work aims to apply the iterative method to solve the discrete analogy of the finite crack and the set of collinear cracks.

The rest of the paper is structured as follows. In \sect{chap:problem}, the problem is formulated for the single crack and the generalised case of the set of collinear cracks in the square lattice. The matrix Wiener-Hopf equation is obtained for both cases. In \sect{chapt:iter} the iterative method is described and the approximate solution is derived. It is evaluated numerically in \sect{chap:numerical_results} alongside with constructing the rationally approximated functions. Concluding remarks and an appendix concerned with the formulation of the Dirichlet analogue  of the considered problem (\fig{fig:rigid_scheme}) are provided at the end.

%% file: 2_problem.tex
\section{Matrix Wiener-Hopf problem formulation}
\label{chap:problem}
We will start with a general information about the square lattice in the presence of a crack. Further, we will formulate the Wiener-Hopf problem for the simple case of a single finite crack in the lattice and the generalised case of $N$ collinear cracks.

\subsection{Square lattice with a single finite crack}
\label{sect:single finite crack}
We consider a square lattice with a single crack aligned with one of the two
fundamental directions. As illustrated in Fig.1a and without loss of
generality, the nodes of the lattice can be chosen to coincide with $(m, n)
\in \mathbb{Z}^2$, while the set $\Sigma$ of nodes either side of the crack can
be written $\Sigma = \Sigma^0 \cup \Sigma^{- 1}$, where
\begin{align}
  \Sigma^0 &= \{ (m, 0) \in \mathbb{Z}^2, 0 < m < L \} \\
  \Sigma^{-1} &= \{ (m, - 1) \in \mathbb{Z}^2, 0 < m < L \}, 
\end{align}
for some positive integer $L > 1$ that we will call the crack length. We are
interested in the scattering problem of an incident plane wave $u_{m,
n}^{\tmop{in}}$ \equ{eq:incident-wave} impinging on the crack and in finding an expression for the
scattered field $u_{m, n}$ and the total field $u_{m, n}^{\tmop{tot}}$,
related by $u_{m, n}^{\tmop{tot}} = u_{m, n}^{\tmop{in}} + u_{m, n}$. All
these fields must satisfy the governing equation of the lattice known as the
discrete Helmholtz equation:
\begin{eqnarray}
  \Delta u_{m, n} + \Omega^2 u_{m, n} = 0 & \tmop{on} & \mathbb{Z}^2 \setminus
  \Sigma,  \label{eq:helm_discr}
\end{eqnarray}
where $\Omega$ is the so-called {\tmem{lattice frequency}}, normally assumed
to be real positive, and $\Delta$ is the five point discrete Laplace operator
defined by
\begin{eqnarray}
  \Delta u_{m, n} & = & u_{m + 1, n} + u_{m - 1, n} + u_{m, n + 1} + u_{m, n -
  1} - 4 u_{m, n}, 
  \label{eq:5p_Laplacian}
\end{eqnarray}
indicating that each node in $\mathbb{Z}^2 \setminus \Sigma$ is assumed to
only have a direct interaction with its four nearest neighbours. This
governing equation comes from considering that the nodes are particles
of equal mass and that the nearest neighbours interaction are represented by
springs of equal stiffness in the out-of-plane direction, in which case $u_{m,
n}$ represents the out-of-plane displacement of the particles (see e.g. \cite{Sharma2015DiffractionCrack}).
Alternatively, it can be seen as an approximation to the standard continuous
Helmholtz equation via a second-order central difference scheme to approximate
the Laplacian (see e.g. \cite{Martin2006DiscreteLattice}) on a square mesh. In both cases, the Helmholtz
equation is derived from the linear wave equation by assuming that the
time-dependent displacements are all time-harmonic.

The boundary conditions are specified by assuming that the links between the
nodes of $\Sigma^0$ and those of $\Sigma^{- 1}$ are broken. That is that each
node of $\Sigma$ only interacts with three neighbours. As a result, we can
write the boundary conditions as
\begin{eqnarray*}
  \Delta^{\Sigma^0} u_{m, n}^{\tmop{tot}} + \Omega^2 u_{m, n}^{\tmop{tot}} &=& 0
  \text{ for } (m, n) \in \Sigma^0, \\
  \Delta^{\Sigma^{- 1}} u_{m,
  n}^{\tmop{tot}} + \Omega^2 u_{m, n}^{\tmop{tot}} &=& 0 \text{ for } (m, n) \in
  \Sigma^{- 1}, 
\end{eqnarray*}
where the boundary operators $\Delta^{\Sigma^0}$ and $\Delta^{\Sigma^{- 1}}$
are defined by
\begin{eqnarray}
  \Delta^{\Sigma^0} u_{m, n} & = & u_{m + 1, n} + u_{m - 1, n} + u_{m, n + 1}
  - 3 u_{m, n}, \label{eq:0_operator}\\
  \Delta^{\Sigma^{- 1}} u_{m, n} & = & u_{m + 1, n} + u_{m - 1, n} + u_{m, n -
  1} - 3 u_{m, n} . \label{eq:1_operator}
\end{eqnarray}

\begin{remark}
    Note that these boundary conditions make sense when
thinking of the masses linked by springs, but do not coincide with the
numerical approximation of a standard boundary condition used for scattering
problems in the continuous case (e.g. Neumann, Dirichlet or Robin).
\end{remark} 

As mentioned above, we choose $u_{m, n}^{\tmop{in}}$ to be a plane wave, i.e.
it can be written as
\begin{eqnarray}
  u_{m, n}^{\tmop{in}} & = & e^{- i (K_m m + K_n n)}, 
  \label{eq:incident-wave}
\end{eqnarray}
for some real Bloch wavenumbers $K_m$ and $K_n$. It can be useful to introduce
the incident angle $\varphi^{\tmop{in}}$, as well as the scalar $K$, defined
such that
\begin{eqnarray*}
  K^2 = K_n^2 + K_m^2, & K_m = K \cos (\varphi^{\tmop{in}}), & K_n = K \sin
  (\varphi^{\tmop{in}}) .
\end{eqnarray*}
For a given lattice frequency $\Omega$, not all such plane waves can
propagate, and the admissible values of $K_m$ and $K_n$ are given by the
{\tmem{dispersion relation}} of the lattice, obtained by inputting
(\ref{eq:incident-wave}) into (\ref{eq:helm_discr}):
\begin{eqnarray}
  \Omega^2 & = & 4 - 2 \cos (K_m) - 2 \cos (K_n) . \label{eq:disp_rel}
\end{eqnarray}
Due to the periodicity of $\cos$, we can restrict our choice of $(K_m, K_n)$
to the so-called Brillouin zone $[- \pi, \pi]^2$. Since $K_m,\ K_n$ are real, it follows from the
dispersion relation that $\Omega$ can only take values within $\left[ 0, 2
\sqrt{2} \right]$ for the lattice to support plane wave propagation. Moreover,
as explained in e.g. \cite{Martin2006DiscreteLattice}, \cite{Vanel2016AsymptoticsFunctions} or \cite{Shanin2024AStructures}, the values $\Omega = 0$, $\Omega = 2$ and
$\Omega = 2 \sqrt{2}$ are resonant degeneracies, and we will omit these values
in the present work. We will hence assume that $\Omega \in \left( 0, 2
\sqrt{2} \right) \setminus \{ 2 \}$.

Upon introducing the auxiliary functions
\begin{eqnarray}
  v_m^{\tmop{in}} \assign u_{m, - 1}^{\tmop{in}} - u_{m, 0}^{\tmop{in}}, &
  \tmop{and} & v_m \assign u_{m, - 1} - u_{m, 0},  \label{eq:vm}
\end{eqnarray}
the boundary problem can be written in terms of the scattered field $u_{m, n}$
as
\begin{equation}
  \left\{ \begin{array}{lllll}
    \Delta u_{m, n} + \Omega^2 u_{m, n} & = & 0 & \tmop{for} & (m, n) \in
    \mathbb{Z}^2 \setminus \Sigma\\
    \Delta^{\Sigma^0} u_{m, n} + \Omega^2 u_{m, n}  & = & v_m^{\tmop{in}} &
    \tmop{for} & (m, n) \in \Sigma^0\\
    \Delta^{\Sigma^{- 1}} u_{m, n} + \Omega^2 u_{m, n}  & = & -
    v_m^{\tmop{in}} & \tmop{for} & (m, n) \in \Sigma^{- 1}
  \end{array} \right. \label{eq:boundary_problem}
\end{equation}
For scattering problems in infinite domains, for the problem
to be well posed, one must specify a condition at infinity: the
{\tmem{radiation condition}}. We will do this through the {\tmem{limiting
absorption principle}}: if $\Omega$ is assumed to have a small positive
imaginary part $\varepsilon_{\Omega}$, we set the condition by imposing that the
resulting scattered field $u_{m, n}$ must be exponentially decaying at
infinity. The true physical solution is then understood as the limit of these
admissible solutions when $\varepsilon_{\Omega}\rightarrow 0^+$.

\subsubsection{Symmetry of the problem}
To analyse the symmetry properties of the problem, we will write $u_{m,n}^{\text{in}}$ as a sum of symmetric $u_{m,n}^{\text{in}; s}$ and anti-symmetric $u_{m,n}^{\text{in}; a}$ components:
\begin{equation}
    u_{m,n}^{\text{in}} = u_{m,n}^{\text{in}; s} + u_{m,n}^{\text{in}; a},
\end{equation}
where
\begin{align}
    u_{m,n}^{\text{in}; s} &= \frac{1}{2}(u_{m,n}^{\text{in}}+u_{m,-1-n}^{\text{in}}), \label{eq:u_in_sym}
    \\
    u_{m,n}^{\text{in}; a} &= \frac{1}{2}(u_{m,n}^{\text{in}}-u_{m,-1-n}^{\text{in}}). \label{eq:u_in_asym}
\end{align}
Then we can split the original problem for $u_{m,n}$ into two, for symmetric and anti-symmetric solutions:
\begin{alignat*}{2}
    u_{m,n}^{s} &= \frac{1}{2}(u_{m,n}+u_{m,-1-n}) &&= u_{m,-1-n}^{s},
    \\
    u_{m,n}^{a} &= \frac{1}{2}(u_{m,n}-u_{m,-1-n}) &&= -u_{m,-1-n}^{a}.
\end{alignat*}

From \equ{eq:vm}, \equ{eq:u_in_sym},\equ{eq:u_in_asym} it follows that:
\begin{align}
    v_m^{\text{in}; s} &= 0,\label{eq:vm_in_sym}
    \\
    v_m^{\text{in}; a} &= v_m^{\text{in}}. \label{eq:eq:vm_in_asym}
\end{align}
Hence, from \equ{eq:boundary_problem} and \equ{eq:vm_in_sym} it follows that there is a trivial solution for the symmetric problem $u_{m,n}^{s}\equiv0$. Due to the uniqueness of the diffraction problem solution, we can consider the symmetric problem trivial. Therefore $u_{m,n} = u_{m,n}^{a}$, and we can formulate the property of anti-symmetry for the solution:
\begin{equation}
    u_{m,n}=-u_{m,-1-n}.
    \label{eq:field_crack_symmetry}
\end{equation}
or, correspondingly, using the definition \equ{eq:vm} of $v_m$, that
\begin{equation}
    v_m = -2u_{m,0}.
    \label{eq:field_crack_symmetry_vm}
\end{equation}
The statement \equ{eq:field_crack_symmetry}
can also be confirmed by the application of the discrete Fourier transform in a similar way to \cite{Sharma2015DiffractionCrack}.
Therefore, instead of solving the problem for all $(m,n)\in\mathbb{Z}^2$ we can now only find a solution for the upper half-plane $n\ge0$.

\subsubsection{The matrix Wiener-Hopf equation}
\label{sect:WH_finite_crack}
Let us denote by $U_n(z)$ the discrete Fourier transform of $u_{m,n}$ in the $m$ direction. That is we have 
\begin{equation}
    U_n(z)=\sum_{m=-\infty}^{\infty} u_{m,n} z^{-m}
    \label{eq:discr_fourier}
\end{equation}

It can be shown (see \ref{app:appB} for details) that $U_n(z)$ \equ{eq:discr_fourier} is analytic in the annulus of analyticity:
\begin{equation}
    \mathcal{A}= \{e^{-\varepsilon}<|z| < e^{\varepsilon}\}. \label{eq:annulus_crack}
\end{equation}
Where $\varepsilon$ is related to the small positive imaginary part of $\Omega$ and therefore $K$ from \equ{eq:disp_rel}. In the limit $\varepsilon\to 0^+ $, $\mathcal{A}$ reduces to the unit circle. We will further refer to the regions of analyticity for functions with `$-$' and `$+$' sub-/superscript as `inside' and `outside' of the unit circle in the $z$ complex plane, and refer to these functions as `minus' and `plus' functions respectively. 

Applying the discrete Fourier transform \equ{eq:discr_fourier}
to \equ{eq:helm_discr}, we have that for all $n\neq 0,-1$:
 \begin{equation}
      U_{n}(z)\{z+z^{-1}-4+\Omega^2\}+U_{n+1}(z)+U_{n-1}(z)=0, \quad z\in \mathcal{A}.
      \label{eq:fourier_full_lattice}
 \end{equation}
For the case of a real $\Omega$ equation \equ{eq:fourier_full_lattice} is valid on the unit circle $|z|=1$, excluding a finite number of isolated special points. 
Details of the general solution of \equ{eq:fourier_full_lattice} for the crack in the square lattice are provided in \cite{Sharma2015DiffractionCrack}, but can be summarised as follows. Because of the radiation condition, $U_n$ decays as $n\to\pm\infty)$, so the general solution of \equ{eq:fourier_full_lattice} in the upper half plane $(n\ge0)$ is given by \cite{Slepyan2002ModelsMechanics}:
\begin{equation}
    U_{n}(z)=U(z)\left(\lambda(z)\right)^n,
    \quad z\in\mathcal{A}
    \label{eq:discr_crack_sol_1}
\end{equation}
where $U(z)$ is an unknown function.

The factor $\lambda(z)$ found to be
\begin{equation}
    \lambda(z) = \frac{1-\mathcal{K}(z)}{1+\mathcal{K}(z)}, 
    \label{eq:discr_lamda_L}
\end{equation}
where 
\begin{equation}
\mathcal{K}(z)=\sqrt{\frac{\mathcal{H}(z)}{\mathcal{R}(z)}}
, \label{eq:K_H}
\end{equation}
and the functions $\mathcal{H}(z)$ and $\mathcal{R}(z)$ are defined as
\begin{align}
    \mathcal{H}(z)&:=2-z-z^{-1}-\Omega^2, 
    \label{eq:H}\\
    \mathcal{R}(z)&:=\mathcal{H}(z)+4. \label{eq:R}
\end{align}

The physical field $u_{m,n}$ can be obtained from $U_n$ by means of the inverse discrete Fourier transform:
\begin{equation}
    u_{m,n}=\frac{1}{2 \pi i} \oint_{\mathcal{C}_{z}} U_n(z) z^{m-1} d z, \quad m \in \mathbb{Z},
    \label{eq:discr_inv_fourier}
\end{equation}
where $\mathcal{C}_{z}$ denotes any rectifiable, closed, counterclockwise contour, which lies in the annulus of analyticity $\mathcal{A}$ of $U_n(z)$ and encircles the origin $z=0$.

Due to the geometry of the problem, we need to write two Wiener-Hopf equations with respect to both crack edges. To do that, we will  introduce two functions $U^{(1)}$ and $U^{(2)}$ defined by:
define partial discrete Fourier transforms of the sequence $u_{m,0}$ in $m$ as follows:
\begin{alignat}{3}  
 U^{(1)}(z)&=\sum_{m=-\infty}^{\infty} u_{m,0} z^{-m}&&=U(z) \label{eq:transf_U1}
 \\
 U^{(2)}(z)&=\sum_{m=-\infty}^{\infty} u_{m,0} z^{-m+L}&&=z^{L}U(z)
\end{alignat}
where $L$ is the length of a crack. 

An important observation for the connection between $U^{(2)}$ and $U^{(1)}$ is: $U^{(2)}=z^L U^{(1)}$. We also define
\begin{align}
    U_{-}^{(1)}&:=\sum_{m=-\infty}^0 u_{m,0} z^{-m}, \label{eq:u1}\\
    U_+^{(2)}&:= \sum_{m=L}^{\infty} u_{m,0} z^{-m+L}. \label{eq:u2}
\end{align}
Note that functions with the subscript `$-$'
are convergent for $|z|\leq e^{\varepsilon}$ and functions with `$+$' are convergent for $|z|\ge e^{-\varepsilon}$, i.e. their common annulus of analyticity is $\mathcal{A}$ \equ{eq:annulus_crack}. 
Moreover, let us define the finite power series function:
\begin{equation}
    U^{(1)}_+ (z)=\sum_{m=1}^{L-1} u_{m,0} z^{-m}.
  \label{eq:u_d}
\end{equation}
So that, from the definition \equ{eq:transf_U1} of $U$ and \equ{eq:u1},\equ{eq:u2},\equ{eq:u_d} we can decompose $U$ as: 
\begin{equation}
    U(z) = U^{(1)}_-(z)+U^{(1)}_+(z)+z^{-L} U_{+}^{(2)}(z). \label{eq:U_decomp_subs}
\end{equation}

From \equ{eq:boundary_problem},\equ{eq:field_crack_symmetry_vm} and \equ{eq:boundary_problem} for $n=0$:
\begin{equation}
    \begin{cases}
        \Delta^{\Sigma^0} u_{m,0} + \Omega^2u_{m,0} = 
        2u_{m,0}, 
        \quad &(m,0)\notin\Sigma^0,\\
        \Delta^{\Sigma^0} u_{m,0} + \Omega^2u_{m,0} = v_m^{\text{in}}, \quad &(m,0)\in\Sigma^0.
    \end{cases}
    \label{eq:n0}
\end{equation}

Applying the discrete Fourier transform to \equ{eq:n0} provides the following equation
\begin{equation}
 z U+z^{-1} U+U_1-3 U + \Omega^2 U = 2U_{-}^{(1)} + 2U_{+}^{(2)} z^{-L} + V_+^{\text{in}},
 \label{eq:subs0_above}
\end{equation}
where 
\begin{equation}
    V_+^{\text{in}}(z) = \sum_{m=1}^{L-1}v_m^{\text{in}} z^{-m}.
\end{equation}
From the general solution \equ{eq:discr_crack_sol_1} we have $U_1 = U\lambda$, therefore \equ{eq:subs0_above} can be rewritten as:
\begin{equation}
    (\mathcal{H}+1-\lambda) U = 2\left[z^{-L} U_{+}^{(2)} + U_{-}^{(1)}\right] - V_+^{\text{in}}. 
    \label{eq:stepp}
\end{equation}
where $\mathcal{H}(z)$ was defined in \equ{eq:H}.

From the definition of $\K$ \equ{eq:K_H} and $\mathcal{R}$ \equ{eq:R} we can express $\mathcal{H}=4\mathcal{K}^2 / (1-\mathcal{K}^2)$.
Then the multiplicative term in front of $U$ in \equ{eq:stepp} can be written as:
\begin{equation}
    \mathcal{H} +1-\lambda  = \frac{4\mathcal{K}^2 }{1-\mathcal{K}^2 } +1 - \frac{1-\mathcal{K} }{1+\mathcal{K} }=\frac{2\mathcal{K} }{1-\mathcal{K} }.
    \label{eq:H simplify}
\end{equation}
Using \equ{eq:H simplify} and \equ{eq:U_decomp_subs}, we can simplify \equ{eq:stepp}  as:
\begin{equation}
z^{-L} U_+^{(2)} +U_{-}^{(1)}+\mathcal{K}U^{(1)}_+ =- \frac{1}{2} V_+^{\text{in}}(1-\mathcal{K}).
\label{eq:crack_eq}
\end{equation}

Note that \equ{eq:crack_eq} is not a Wiener-Hopf equation yet, since it contains three unknown functions $U_{-}^{(1)}$, $U^{(1)}_+ $ and $U_+^{(2)} $. 
To get an equation of the desired form let us observe the properties of analyticity of $U^{(1)}_+ $. Note that $U^{(1)}_+ $
is a `plus' function, analytic outside the unit circle. However, if we multiply \equ{eq:u_d} by $z^{L}$
\begin{equation}
	z^LU^{(1)}_+ =\sum_{m=1}^{L-1}u_{m,0}z^{L-m} =: U^{(2)}_-
\label{eq:trick}
\end{equation}
the result turns into a `minus' function, analytic inside the unit circle. Hence the following system of 2 equations for 4 unknown functions is obtained:
\begin{align*}
\begin{cases}
U_{-}^{(1)} + z^{-L} U_{+}^{(2)} +\mathcal{K}U^{(1)}_+ &=-\frac{1}{2} V_+^{\text{in}}(1-\mathcal{K}) \\
U^{(2)}_- - z^LU^{(1)}_+  &=0
\end{cases}.
\end{align*}

The final matrix Wiener-Hopf equation with a triangular matrix kernel is then:
\begin{equation}
\left(\begin{array}{cc}
\mathcal{K} & z^{-L} \\ \\
-z^L & 0
\end{array}\right)
\left(\begin{array}{c}
U^{(1)}_+  \\ \\
U_{+}^{(2)} 
\end{array}\right) 
=
-
\begin{pmatrix}
U_{-}^{(1)} \\ \\
U^{(2)}_- 
\end{pmatrix}
+
\begin{pmatrix} 
f(z) \\ \\
0
\end{pmatrix}, 
\label{eq:matrixWH_crack}
\end{equation}
\begin{equation}
    \text{where} \quad f(z) = \frac{1}{2} V_+^{\text{in}}(\mathcal{K}-1). \label{eq:f_z}
\end{equation}
Note that the top-left entry of the matrix is the kernel for the semi-infinite crack problem \cite{Sharma2015DiffractionCrack}.
The forcing term \equ{eq:f_z} can be simplified as follows:
\begin{multline}
    f(z) = (\mathcal{K}-1) \frac{1}{2} \sum_{m=1}^{L-1} \left(e^{-imK_m +iK_n}- e^{-imK_m}\right) z^{-m} \\
    = (\mathcal{K}-1) \frac{e^{iK_n}-1}{2} \sum_{m=1}^{L-1} \left(\frac{z_{\text{p}}}{z}\right)^m, \label{eq:fz_series}
\end{multline}
where $z_{\text{p}}:= e^{-iK_m}$ is a constant, defined by the incident field angle and lattice wave number.

The kernel of the matrix equation \equ{eq:matrixWH_crack} is a triangular $2\times2$ square matrix, which contains factors $z^L$ and $z^{-L}$, that are infinitely growing as $z$ approaches $\infty$ or $0$ respectively. In a continuous analogue of the problem of diffraction by a finite strip \cite{Kisil2018AerodynamicExtensions}, these factors are replaced by $e^{\pm iL\alpha}$, which are growing exponentially at upper/lower half-planes as $\alpha\to\pm\infty$. Currently, constructive methods to factorise such matrices do not exist.
However, in the discrete case, all the entries of the matrix kernel in \equ{eq:matrixWH_crack} have at most algebraic growth at infinity. Therefore it technically can be factorised directly, for example, by applying an asymptotic method from \cite{Mishuris2014AnFunctions}. Moreover, it was shown in \cite{Sharma2020ScatteringZone} that a matrix equation similar to \equ{eq:matrixWH_crack} can be reduced to a finite system of linear equations. For both these methods, the computational cost grows with the length of the crack,  and for the case of $N$ cracks,  the problem becomes significantly more complicated. The iterative method, discussed in this work, on the opposite, converges faster in the case of a long crack when $L\gg1$.

\begin{remark}
A matrix equation similar to \equ{eq:matrixWH_crack} can also be obtained for a finite rigid constraint on a lattice (see \ref{appx:wh_rigid} for the details). We are not going to further consider that case since it is in all aspects similar to the crack problem.
\end{remark}

\subsubsection{Exact solution in terms of discrete Green's function}
\label{sect:Greens_function}
As mentioned before, the exact solution for the problem at hand can be obtained \cite{Sharma2015Near-tipCrack}. For a crack of length $L$, it reads:
\begin{gather}
    u_{m, n}=-\sum_{j=1}^{L-1}\left(v_j+v_j^{\text{\normalfont in}}\right)\left(\mathcal{G}_{m-j, n}-\mathcal{G}_{m-j, n+1}\right), \label{eq:ex_sol_crack}
\end{gather}
where $\mathcal{G}_{m,n}$ is the lattice Green's function (see e.g. \cite{Martin2006DiscreteLattice} or \cite{Vanel2016AsymptoticsFunctions}):
\begin{gather*}
\mathcal{G}_{m,n}=\frac{1}{4 \pi^2} \int_{-\pi}^\pi \int_{-\pi}^\pi \frac{\cos m \xi \cos n \eta}{\sigma_S\left(\xi, \eta, \Omega^2\right)} d \xi d \eta, \quad(m, n) \in \mathbb{Z}^2,
\label{eq:green_func_def}
\\
\text{where}\quad \sigma_S\left(\xi, \eta, \Omega^2\right)=\Omega^2-4+2\cos{\xi}+2\cos{\eta},\quad (\xi,\eta)\in[-\pi,\pi]^2.
\end{gather*}
Here
$v_j^{\text{\normalfont in}}$ (resp.\ $v_j$), defined in \equ{eq:vm}, relates to the incident wave (resp. unknown scattered field) on the crack nodes. Using \equ{eq:field_crack_symmetry_vm} and evaluating \equ{eq:ex_sol_crack} at $n=0$ for $m=1,...,L-1$ leads to a linear system for the vector $\boldsymbol{v}=[v_1,...,v_{L-1}]^\top$, whose reduction leads to:
$$
\boldsymbol{v}=\left(\mathbf{I}_{L-1}-\boldsymbol{F}_{L-1}\right)^{-1} \boldsymbol{F}_{L-1} \boldsymbol{v}^{\text{\normalfont in}},
$$
where $\mathbf{I}_{L-1}$ is the $(L-1)\times(L-1)$ identity matrix and 
$\boldsymbol{F}_{L-1}$ is the $(L-1)\times(L-1)$ symmetric Toepliz matrix, whose $(i,j)^{\text{th}}$ component is given by $\left[\boldsymbol{F}_{L-1}\right]_{i j}=2 \mathcal{G}_{i-j, 1}-2 \mathcal{G}_{i-j, 0}$. 
As has been mentioned previously, although it is a linear system, each $\left[\boldsymbol{F}_{L-1}\right]_{i j}$ requires the calculation of a discrete Green's function as a double integral. So the computational cost for a large crack length $L$ is high since the dimension of $\boldsymbol{F}_{L-1}$
is growing with $L$. 
Note that the formula \equ{eq:ex_sol_crack} can also be understood as a special case of the boundary algebraic equations method, described in \cite{Poblet-Puig2015SuppressionEquations}. We have implemented this method for comparison in \sect{sect:results_plots}.

\subsection{N collinear finite cracks}
\label{sect:colliear_cracks}
Let us generalise the formulation \equ{eq:matrixWH_crack} to a slightly more complicated scattering problem on a lattice with $N$ collinear cracks (\fig{fig:cracks_scheme}). Below, we will show that this problem can also be reduced to a matrix Wiener-Hopf equation, but with a $2N\times 2N$ matrix kernel. 

For this section, we will define partial discrete Fourier transforms along the row $n=0$ for $\ell\in\{1,N\}$ as follows:
\begin{align}
    U_-^{(1)}(z) &= \sum_{m=-\infty}^{m_1} u_{m,0}\; z^{-m}, \label{eq:minus}\\
    U_+^{(2\ell-1)}(z)  &= \sum_{m=m_{(2\ell-1)}+1}^{m_{2\ell}-1} u_{m,0}\; z^{-m+m_{(2\ell-1)}}, \label{eq:transf_odd}\\
    U_+^{(2\ell)}(z)  &= \sum_{m=m_{2\ell}}^{m_{(2\ell+1)}} u_{m,0}\; z^{-m+m_{2\ell}}, \label{eq:transf_even}
\end{align}
where $m_\ell$ are the horizontal indices for nodes on the cracks edges, $m_{1} =0$, $m_{(2N+1)} =\infty$ (see \fig{fig:cracks_scheme}). Here $U_+^{(2\ell-1)} $ corresponds to the transform along the crack faces, and $U_+^{(2\ell)}$ are the transforms along the nodes with all four links intact.

Hence, as for \equ{eq:crack_eq}, we can derive the following equation for the scattering by $N$ cracks:
\begin{equation}
    U_-^{(1)} + \sum_{\ell=1}^{2N} z^{-m_{\ell}} E_{\ell} U_+^{(\ell)} = f_N,
    \label{eq:gen_sol_mult_cracks}
\end{equation}
where $E_{2\ell}(z)=1$ and $E_{(2\ell-1)}(z)=\mathcal{K}(z)$ for $\ell\in\{1,N\}$ and
\begin{equation}
    f_N(z) = \frac{e^{iK_n}-1}{2} (\mathcal{K}(z)-1) \sum_{\ell=1}^N\left[\sum_{m=m_{(2\ell-1)}+1}^{m_{2\ell}-1} \left(\frac{z_{\text{p}}}{z}\right)^m \right], \quad z_{\text{p}} := e^{-iK_n}. \label{eq:f_mult_series}
\end{equation}

Similarly to \equ{eq:trick}, we can make a note on the analytic properties of \equ{eq:transf_odd} and \equ{eq:transf_even}, by introducing the minus function:
\begin{equation}
    U_-^{(\ell+1)} := z^{m_{(\ell+1)}}U_+^{(\ell)}, \quad \ell\in\{1,2N-1\}.
    \label{eq:analyt_prop_set}
\end{equation}
Hence, a matrix analogue to \equ{eq:matrixWH_crack} is obtained:
\begin{multline}
    \begin{pmatrix} 
    E_1 & E_2 z^{(m_1-m_2)} & \cdots & \cdots & E_{2N} z^{(m_1-m_{2N})} \\
    z^{(m_2-m_1)} & 0 & \cdots  & \cdots & 0 \\
    \vdots  & \vdots &    & 0 & \vdots\\ 
    0 & 0 & \cdots & z^{(m_{2N}-m_{2N-1})} & 0 
    \end{pmatrix} 
    \begin{pmatrix} 
    U_+^{(1)} \\
    U_+^{(2)}\\
    \vdots\\ 
    U_+^{(2N)}
    \end{pmatrix}  
    = \\
    =-\begin{pmatrix} 
    U_-^{(1)} \\
    U_-^{(2)}\\
    \vdots\\ 
    U_-^{(2N)}
    \end{pmatrix} +
    \begin{pmatrix} 
    f_N \\
    0\\
    \vdots\\ 
    0
    \end{pmatrix}.
\label{eq:mult_cracks_triang}
\end{multline}

\begin{remark}
The expressions above are obtained for a set of finite cracks, however, a similar system can be written when the first and/or last crack are semi-infinite. One should be careful to split the transform correctly with respect to the crack edges.
However, in that case, the forcing term \equ{eq:f_mult_series} will have a pole, which would require a more careful integration when taking the inverse discrete Fourier transform, and will lead to geometrically reflected waves in the solution.
\end{remark}

%% file: 3_iterative.tex
\section{The iterative method}
\label{chapt:iter}
The Wiener-Hopf method for solving an equation of the form
$$U_{+}(z)+\mathcal{K}(z)U_{-}(z)=D(z),$$
where $U_{\pm}$ are unknown and $\mathcal{K}(z),\ D(z)$ are known functions, can be summarised in a few steps \cite{Kisil2021TheMethods}:
\begin{enumerate}
    \item Factorise the kernel $\mathcal{K}(z)$:
    \begin{equation}
    \mathcal{K}(z)=\mathcal{K}^+(z)\mathcal{K}^-(z) 
    \end{equation}
    and sum split the forcing term $\widetilde{D}(z):=D(z)/\K^+(z)$:
    \begin{equation}
    \quad\quad \widetilde{D}(z) = \widetilde{D}_-(z)+\widetilde{D}_+(z). 
    \end{equation}
    Here the subscript $\pm$ denotes additive splitting, while multiplicative factors are marked by superscript $\pm$, i.e., for a suitable function $f$:
    \begin{equation}
        f = f^- f^+, \quad f = f_- + f_+.
        \label{eq:subscript_def}
    \end{equation}
    \item Study the behaviour of the known and unknown functions at $0$ and $\infty$. By the generalised Liouville theorem, the `plus' and `minus' parts of the Wiener-Hopf equation are equal to the same polynomial function $J(z)$, which is often zero:
    $$\frac{U_{+}(z)}{\mathcal{K}^+(z)}-\widetilde{D}_+(z)=-\mathcal{K}^-(z) U_{-}(z)+\widetilde{D}_-(z):=J(z).$$
    Hence the two unknowns $U_{\pm}$ can be found separately and the problem is solved.
    \item To get back to the physical unknown, we use the inverse Fourier transform. 
\end{enumerate}

Matrix kernels cannot be factorised in general, but additive splitting is relatively easy.
The iterative method \cite{Kisil2018AnFactors} allows us to avoid the direct factorisation of the matrix kernel and use additive splitting instead.
In this section, we will provide a detailed description of the iterative method applied to the matrix Wiener-Hopf equations that arise from the scattering problems for a set of $N$ collinear cracks in the square lattice (\sect{chap:problem}).

The first step is to partition the kernel in \equ{eq:mult_cracks_triang} into two parts
to place the growing factors in the right place in the equation for them to pre-multiply functions $U_{\pm}$ of the corresponding analyticity. 
To achieve this we will need to re-scale \equ{eq:gen_sol_mult_cracks} by $z^{m_{\ell}}$ for each $\ell\in\{1,N\}$ to provide a set of $2N$ equations. Together they yield a matrix equation
\begin{equation}
    \mathbf{A}\mathbf{U}_- + \mathbf{B}\mathbf{U}_+ = \mathbf{F},
    \label{eq:matrox_mult_cracks_WH}
\end{equation}
where $\mathbf{A}$ and $\mathbf{B}$ are $2N\times2N$ matrices:
\begin{gather*}
    \mathbf{A} = 
    \begin{pmatrix}
        1 & 0 & \cdots & 0\\
        z^{m_{1}} & E_1 &  & \vdots \\
        \vdots & \vdots &\ddots & 0  \\ 
        z^{m_{2N}} &  E_1 z^{m_{(2N-1)}}&\cdots & E_{(2N-1)} 
    \end{pmatrix}, \\
    \mathbf{B} = 
    \begin{pmatrix}
        E_1 & E_2z^{-(m_1-m_{2})}& \cdots & E_{2N} z^{(m_1-m_{2N})} \\
        0 & E_2 & \cdots  & E_{2N} z^{(m_2-m_{2N})} \\ 
        \vdots & \vdots & \ddots & \vdots \\ 
        0 &\cdots &0 & E_{2N} \\      
    \end{pmatrix},
\end{gather*}
 $\mathbf{U}_{\pm} = [U_{\pm}^{(1)}, U_{\pm}^{(2)}, \cdots, U_{\pm}^{(2N)}]^\top$,  and $\mathbf{F} = [F_1, F_2, \cdots, F_{2N}]^\top$ with $F_{\ell}(z) = z^{m_\ell}f_N(z).$

Then from \equ{eq:matrox_mult_cracks_WH} we have $2N$ scalar Wiener-Hopf equations:
\begin{equation}
    \mathbf{A}_{\ell\ell}U_-^{(\ell)}+\mathbf{B}_{\ell\ell}U_+^{(\ell)} = \widetilde{F}_{\ell},\label{eq:scalar_Wiener_Hopf_cracks}
\end{equation}
where the forcing term $\widetilde{F}_l$ is defined as:
\begin{gather}
    \widetilde{F}_{\ell} =F_{\ell} + F^*_{\ell},\quad F^*_{\ell} = - \sum^{2N}_{\substack{{p=1}\\ p\neq \ell}} \mathbf{A}_{\ell p}U_-^{(p)} - \sum^{2N}_{\substack{{p=1}\\ p\neq l}}\mathbf{B}_{\ell p}U_+^{(p)}.
    \label{eq:forcing_term_F}
\end{gather}
Here the subscript $\ell\ell$ or $\ell p$ refer to the indices of the entry of the matrix. Due to the nature of the problem of the set of finite cracks either $\mathbf{A}_{\ell\ell}=1, \ \mathbf{B}_{\ell\ell}=\mathcal{K} $ for odd $\ell=2p-1,\ p\in\{1,N\}$ or $\mathbf{A}_{\ell\ell}=\mathcal{K}, \ \mathbf{B}_{\ell\ell}=1$ for even $\ell=2p,\ p\in\{1,N\}$.

Assuming the forcing term $\widetilde{F}_{\ell}$ is known, we can  rearrange \equ{eq:scalar_Wiener_Hopf_cracks} for odd indices
$\ell=2p-1,\ p\in[1,N]$ as:
\begin{equation}
    - \frac{U_-^{(\ell)}}{\mathcal{K}^-} 
    +
    \left(\frac{\widetilde{F}_{\ell}}{\mathcal{K}^-}\right)_- 
    = 
    \mathcal{K}^+U_+^{(\ell)} 
    -
    \left(\frac{\widetilde{F}_{\ell}}{\mathcal{K}^-}\right)_+ ;
    \label{eq:scalar_Wiener_Hopf_cracks_odd}
\end{equation}
and for even indices $\ell=2p,\ p\in[1,N]$:
\begin{equation}
    -\mathcal{K}^-U_-^{(\ell)} 
    +
    \left(\frac{\widetilde{F}_{\ell}}{\mathcal{K}^+}\right)_- 
    =
    \frac{U_+^{(\ell)}}{\mathcal{K}^+} 
    -
    \left(\frac{\widetilde{F}_{\ell}}{\mathcal{K}^+}\right)_+.
    \label{eq:scalar_Wiener_Hopf_cracks_even}
\end{equation}

The left parts in \equ{eq:scalar_Wiener_Hopf_cracks_odd} and \equ{eq:scalar_Wiener_Hopf_cracks_even} are analytic inside the unit circle and the right parts are analytic outside of the unit circle. Thus, by application of Liouville's theorem, the left and right parts of \equ{eq:scalar_Wiener_Hopf_cracks_odd}-\equ{eq:scalar_Wiener_Hopf_cracks_even} are equal to the same polynomials $J_1^{(\ell)}(z)$ and $\ J_2^{(\ell)}(z)$ respectively. As it will be discussed further in \sect{sect:additive_splitting}, additive splitting is defined in such a way that for any function $G$: $G_+(z) \to 0$ as $z\to\infty$, $G_-(z) \to C$ as $z\to0$, where $C:=G_-(0)$ is a constant. 
This, along with the definition of $U_\pm^{(\ell)}$ \equ{eq:minus}-\equ{eq:transf_even} guarantees that $J_{1,2}^{(\ell)}(z)$ are actually constant. Let us define the constants
\begin{align}
    C_1^{(\ell)} = \left(\frac{\widetilde{F}_{\ell}}{\mathcal{K}^-}\right)_-(z=0), \label{eq:contant1}\\
    C_2^{(\ell)} = \left(\frac{\widetilde{F}_{\ell}}{\mathcal{K}^+}\right)_-(z=0). \label{eq:contant2}
\end{align}

It can also be shown from the definition \equ{eq:K_H} that:
\begin{equation}
    \begin{array}{ccc}
         \mathcal{K}^-\to C_3 & \text{as} & z\to0,  \\
         \mathcal{K}^+\to C_4 & \text{as} & z\to\infty, 
    \end{array}
\end{equation}
where $C_{3,4}$ are constants, related by $C_3C_4=1$. Moreover, from \equ{eq:minus},\equ{eq:analyt_prop_set}
\begin{equation*}
    \begin{array}{cccc}
         &U_-^{(1)}\to u_{0,0} &\text{as} &z\to0,  \\
         \forall \ell\in\{2,2N\} & U_-^{(\ell)}\to 0  & \text{as} & z\to0, 
    \end{array}
\end{equation*}
from \equ{eq:transf_odd} and \equ{eq:transf_even}
\begin{equation*}
    \begin{array}{ccccc}
         \text{for}& \ell=2p-1,\ p\in\{1,N\} &U_+^{(\ell)}\to 0   &\text{as} & z\to\infty,  \\
         \text{for}&  \ell=2p,\ p\in\{1,N\} & U_+^{(\ell)}\to u_{m_{\ell},0}  & \text{as} & z\to\infty, 
    \end{array}
\end{equation*}
where $u_{m_{\ell},0}$ is an unknown constant. Thus, taking limits for the left and right parts of \equ{eq:scalar_Wiener_Hopf_cracks_odd} ,\equ{eq:scalar_Wiener_Hopf_cracks_even} we can show that for $p\in\{2,N\}$:
\begin{equation}
\begin{array}{lllllll}
    J_1^{(1)} &=& -u_{0,0}\times C_3^{-1}  + C_1^{(1)} &=& 0\times C_4 - 0 &:=& 0, \\
    J_2^{(2)} &=& -0 \times C_3 + C_2^{(2)} &=& u_{m_2,0} \times C_4^{-1} - 0 &:=& C_2^{(2)},\\
    J_1^{(2p-1)} &=& -0\times C_3^{-1} + C_1^{(2p-1)} &=& 0\times C_4 - 0 &:=& 0, \\
    J_2^{(2p)} &=& -0\times C_3 + C_2^{(2p)} &=& u_{m_{(2p)},0}\times C_4^{-1} - 0 &:=& C_2^{(2p)}.
\end{array}
\end{equation}

Thus, for odd indices
$\ell=2p-1,\ p\in\{1,N\}$:
\begin{align}
    U_+^{(\ell)} = \frac{1}{\mathcal{K}^+}\left(\frac{\widetilde{F}_\ell}{\mathcal{K}^-}\right)_+  &= \frac{1}{\mathcal{K}^+}\left(\frac{\left(\widetilde{F}_{\ell}\right)_+}{\mathcal{K}^-}\right)_+ ;
    \label{eq:U_plus_odd}\\
    U_-^{(\ell)} = \mathcal{K}^-\left(\frac{\widetilde{F}_\ell}{\mathcal{K}^-}\right)_- &= \left(\widetilde{F}_{\ell}\right)_- + \mathcal{K}^-\left(\frac{\left(\widetilde{F}_{\ell}\right)_+}{\mathcal{K}^-}\right)_-; \label{eq:U_minus_odd}
\end{align}
and for even indices $\ell=2p,\ p\in[1,N]$:
\begin{align}
    U_+^{(\ell)} = \mathcal{K}^+\left[\left(\frac{\widetilde{F}_\ell}{\mathcal{K}^+}\right)_+ + C_2^{(\ell)}\right]
    &= 
    \left(\widetilde{F}_{\ell}\right)_+ + \mathcal{K}^+\left[\left(\frac{\left(\widetilde{F}_{\ell}\right)_-}{\mathcal{K}^+}\right)_++ C_2^{(\ell)}\right];
    \label{eq:U_plus_even} \\
    U_-^{(\ell)} = \frac{1}{\mathcal{K}^-}\left[\left(\frac{\widetilde{F}_\ell}{\mathcal{K}^+}\right)_- - C_2^{(\ell)}\right]  &= \frac{1}{\mathcal{K}^-}\left[\left(\frac{\left(\widetilde{F}_{\ell}\right)_-}{\mathcal{K}^+}\right)_- - C_2^{(\ell)}\right];
    \label{eq:U_minus_even} 
\end{align}

$C_2^{(\ell)}$ in \equ{eq:U_plus_even}-\equ{eq:U_minus_even} can be found from \equ{eq:contant2}. An exact solution for each $\ell$ requires the knowledge of the functions $U_{\pm}^{(p)}, \  p\in[1,2N]\backslash \{\ell\}$ in the forcing term $\widetilde{F}_{\ell}$ \equ{eq:forcing_term_F}. So in order to move forward it is required to make some approximation. On the first step
we assume that
\begin{gather*}
    \left\{U_{\pm}^{(\ell)}\right\}^{1} = 0, \  \ell\in\{2,2N\},\\
    \text{then}\quad \left\{ F_1^{*} \right\}^{1} = 0, \quad \left\{ \widetilde{F}_1\right\}^{1} = F_1,
\end{gather*}
where the superscript $\{ \}^1$ indicates the first iteration. The physical meaning of this is that we neglect the impacts of each crack edge except the first one. 
Therefore we can find $\{U_{\pm}^{(1)}\}^1$ from \equ{eq:U_plus_odd}-\equ{eq:U_minus_odd}. Then we can substitute the solution for $\{U_{\pm}^{(1)}\}^1$ into \equ{eq:forcing_term_F}, while still assuming that $\{U_{\pm}^{(\ell)}\}^1=0$, $\ell\in\{2,2N\}$. Then the forcing term is
\begin{equation*}
    \left\{  F_2^{*}  \right\}^{1} = - \mathbf{A}_{21}\left\{  U_-^{(1)} \right\}^{1} - \mathbf{B}_{21}\left\{  U_+^{(1)} \right\}^{1}
\end{equation*}
and we can find $\{U_{\pm}^{(2)}\}^1=0$ \equ{eq:U_plus_even}-\equ{eq:U_minus_even}. Physically that means that we considered interaction between the first two crack edges. We now can substitute the first four known functions into \equ{eq:forcing_term_F} for $\ell=3$ and find $\{U_{\pm}^{(3)}\}^1=0$. Subsequently repeating this procedure we reach $\ell=2N$, for which each term in \equ{eq:forcing_term_F} is known:
\begin{gather*}
    \left\{  F_{2N} ^{*}  \right\}^{1} = - \sum^{2N-1}_{p=1}\mathbf{A}_{2N p}\left\{  U_-^{(p)} \right\}^{1} - \sum^{2N-1}_{p=1}\mathbf{B}_{2N p}\left\{  U_+^{(p)} \right\}^{1}.
\end{gather*}
Therefore we solved \equ{eq:scalar_Wiener_Hopf_cracks} for $\{U_{\pm}^{(\ell)}\}^1$, $\ell\in\{2,2N\}$. 
Hence for the $2N^{th}$ equation, we would consider scattering on all the crack edges in the forcing term, coupling the first and the last crack edge interaction. 

To perform the full coupling we now need to start the next iteration, using the known functions from the first loop $\{U_{\pm}^{(l)}\}^1$ as the known data in \equ{eq:forcing_term_F}  and we repeat the same procedure again. The number of iterations required will be discussed later in \sect{chap:numerical_results}, but in realistic problems 3-5 is normally enough. 
Technically, within one iteration, we can solve the equations in any possible order, however, two following iteration strategies seem to be the most reasonable: solving equations in a repeated pattern `forward-forward', when we solve equations in order $1^{st},\ 2^{nd}, ... 2N^{th}$, or `forward-backwards', when the order of equations is circular: $1^{st},\ 2^{nd}, ...,\ 2N^{th}, (2N-1)^{th},...,1^{st}$. The latter can demonstrate a faster convergence \cite{Priddin2020ApplyingFactors} due to the fact that the last crack edge is interacting with the previous one stronger than with the first edge, so the `forward-backwards' strategy appears to be justified by the physics of the process. 

The iterative formula for moving forward which is implemented for the numerical calculation in \sect{chap:numerical_results} is:
\begin{multline}
    \left\{  F^{*}_{\ell} \right\}^{j} = - \sum^{2N}_{p=\ell+1} \left(\mathbf{A}_{\ell p}\left\{  U_-^{(p)} \right\}^{j-1}+\mathbf{B}_{\ell p}\left\{  U_+^{(p)} \right\}^{j-1}\right) \\
    - \sum^{\ell-1}_{p=1}\left(\mathbf{A}_{\ell p}\left\{  U_-^{(p)} \right\}^{j} +\mathbf{B}_{ \ell p}\left\{  U_+^{(p)} \right\}^{j}\right). \label{eq:forward_f}
\end{multline}

%% file: 4_results.tex
\section{Numerical evaluation}
\label{chap:numerical_results}
In this chapter, we will provide details on the numerical implementation of the algorithm described in \sect{chapt:iter} via \verb+Matlab+ and deliver numerical results such as heatmaps and convergence plots for the single crack problem as well as two collinear cracks.

\subsection{Rational approximation}
The additive splitting or factorisation of functions of arbitrary form can be challenging and requires the use of Cauchy-type integrals, which are not easy for practical implementation. A way to simplify this step is to construct a rational approximation of each matrix entry using the \verb+AAA+ algorithm \cite{Nakatsukasa2018TheApproximation} from the \verb+Chebfun+ toolbox in \verb+Matlab+, this method has previously been used in  \cite{Nethercote2022DiffractionScatterers}. Then the factorisation and additive splitting of the required functions on each iterative step become straightforward.

\subsection{Factorisation}
Note that the function $\mathcal{K}$ is the only non-rational entry of the matrix kernel \equ{eq:mult_cracks_triang}. First, we need to get a highly accurate rational approximation for this function of the form:
\begin{equation}
    \mathcal{K}(z)\simeq\widetilde{\K}(z)= K_1\prod_{l=1}^{\mathcal{N}}\frac{z-z_0^l}{z-z_p^l},
    \label{eq:K_rat_fact}
\end{equation}
where $z_0^l$ are the zeros of $\widetilde{\K}$ and $z_p^l$ are its poles. Note that $\forall l,r: l\neq r$ $z_0^l\neq z_0^r,\ z_p^l\neq z_p^r$, i.e.\ all the poles and zeros are simple.

\begin{figure}[H]
\vfill
\centering
    \includegraphics[width=0.6\textwidth]{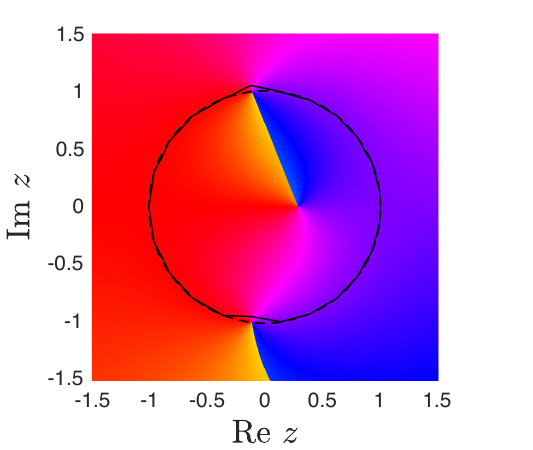} 
    \caption{Phase plot for $\K(z)$. The solid black line corresponds to the integration contour $\mathcal{C}_z$ and the dashed line is the unit circle $|z=1|$. }
    \label{fig:curved_contour}
\end{figure}

It is required to obtain an approximation of the function $\K(z)$ on the integration contour $\mathcal{C}_z$ for the inverse transform \equ{eq:discr_inv_fourier}. For the limiting case when the imaginary part of $\Omega$ tends to zero $(\varepsilon\to0)$, the branch points of $\K(z)$ hit the unit circle, so we have to deform the integration contour to avoid these special points (\fig{fig:curved_contour}). Due to the highly oscillatory nature of the integrand function, we would like to preserve the unit circle shape as much as possible. Therefore, $\mathcal{C}_z$ is defined as a circle\footnote{For the purpose of using `Waypoints' for integration in \verb+Matlab+, in practice the contour is evaluated as a piecewise linear contour, the `corners' of which are lying on the unit circle, except those in front of the special points (see \fig{fig:curved_contour})}
with two small indentations against the special points. 

Note that $\mathcal{K}$ is symmetric with respect to the unit circle, i.e. $\K(z)=\K(1/z)$, and we would like to preserve this symmetry as much as possible while doing rational approximation. To achieve that, we will make a change of variables $\alpha = z+z^{-1}$ and then find the rational approximation of the function approximate the function $\Hat{\K}(\alpha)=\Hat{\K}(z+z^{-1})$, which is equal to $\K(z)$ on the contour $\mathcal{C}_z$. Then each pole (zero) of $\Hat{\K}(\alpha)$ produces two poles (zeros) for $\K(z)$ as $z=\alpha/2\pm\sqrt{\alpha^2-4}/2$.

\begin{figure}[H]
    \centering
     \begin{subfigure}[b]{0.4\textwidth}
         \centering
         \caption{$\widetilde{\K}^+(z)$}
         \includegraphics[width=\textwidth]{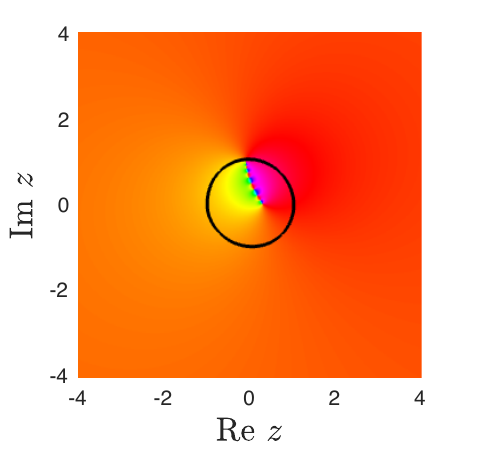} 
         \label{fig:fact_kernel3}
     \end{subfigure}
     \begin{subfigure}[b]{0.4\textwidth}
         \centering
         \caption{$\widetilde{\K}^-(z)$}
        \includegraphics[width=\textwidth]{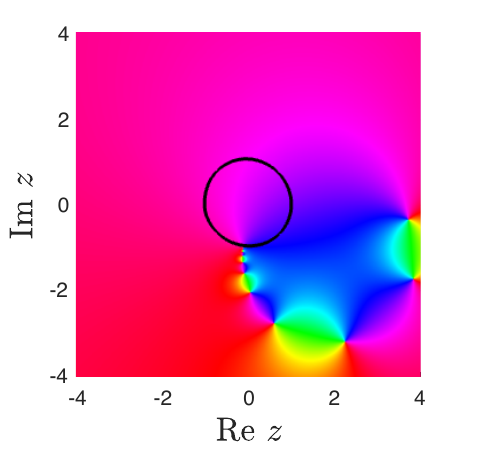} 
         \label{fig:fact_kernel4}
     \end{subfigure}
     \begin{subfigure}[b]{0.4\textwidth}
         \centering
         \caption{$\widetilde{\K}^+(z)\widetilde{\K}^-(z)$}
         \includegraphics[width=\textwidth]{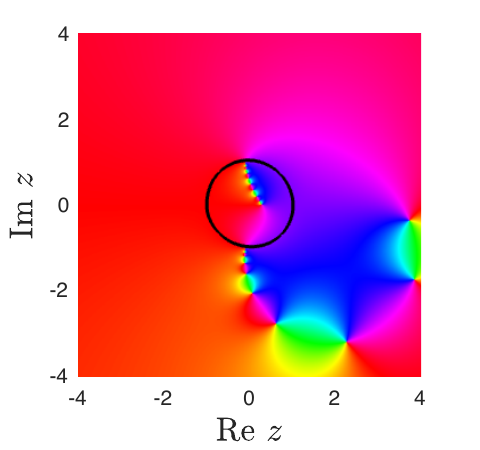} 
         \label{fig:fact_kernel2}
     \end{subfigure}
     \begin{subfigure}[b]{0.4\textwidth}
         \centering
        \caption{$\K(z)$}
         \includegraphics[width=\textwidth]{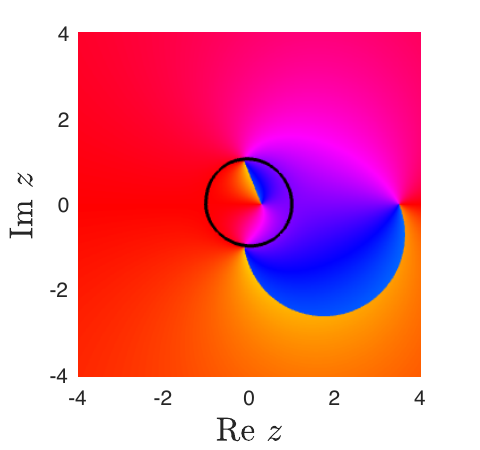}
         \label{fig:fact_kernel1}
     \end{subfigure}
     \caption{Phase plots for the original and rationally aproximated function $\K$ and its factors. On (a)-(c) points denote zeros when colours are ordered from red to blue in the anti-clockwise direction and poles for the clockwise direction. Here the sequence of zero-pole pairs imitates the branching behaviour of the original function. The colour symbolises the phase of a plotted complex function. }
\label{fig:fact_kernel}
\end{figure}

In the next step, we make a change of variables back to $\mathcal{K}(z)$ and factorise the function as follows:
\begin{equation*}
    \widetilde{\K} = \widetilde{\K}^+\widetilde{\K}^- = K_1\prod_{l=1}^{\mathcal{N}}\frac{z-z_0^{-l}}{z-z_p^{-l}}\prod_{l=1}^{\mathcal{N}}\frac{z-z_0^{+l}}{z-z_p^{+l}},
\end{equation*}
where $z_0^{-l},\ z_p^{-l}$ are zeros and poles inside the unit circle (i.e. in the `minus' region) and $z_0^{+l},\ z_p^{+l}$ are outside of the unit circle. Note that due to the trick with the change of variables we have preserved the original symmetry of $\K$. As such, the zeros and poles are now in reciprocal pairs and now satisfy the following relation:
$$z_0^{-l} = \frac{1}{z_0^{+l}}, \quad z_p^{-l} = \frac{1}{z_p^{+l}},$$
and therefore 
$$\widetilde{\K}^+(z)=\widetilde{\K}^-(z^{-1}).$$
The result of the factorisation is presented in \fig{fig:fact_kernel}. We can see indeed that `plus' (\fig{fig:fact_kernel3}) and `minus' (\fig{fig:fact_kernel4}) factors are pole- and zero-free in the regions of their analyticity, and that their product (\fig{fig:fact_kernel2}) returns the rational approximation of the original function (\fig{fig:fact_kernel1}). The accuracy of the approximation taken in this work is $\sim 10^{-7}$ and can be increased by adding more poles and zeros (i.e. increasing $\mathcal{N}$ in \equ{eq:K_rat_fact}) within the \verb+AAA+ algorithm.

\subsection{Additive splitting}
\label{sect:additive_splitting}
When all the entry functions are in a rational form, additive splitting of such functions becomes trivial. If $f(z)$ only has $P$ simple poles $z_{p}^{l},\ l\in\{1,P\}$, we can decompose it as a partial fraction:
\begin{equation}
f(z) = \frac{b(z)}{a(z)}=\frac{r_{P}}{z-z_{p}^{P}}+\frac{r_{P-1}}{z-z_{p}^{(P-1)}}+\ldots+\frac{r_1}{z-z_{p}^1}+r_0(z), \label{eq:partfrac}    
\end{equation}
where the polynomial $r_0(z)$ appears when $b(z)$ is of a higher or equal order than $a(z)$ (then $f(z)$ is a \textit{non-proper rational function}), $r_0(z)$ is a constant if $a$ and $b$ are of the same order. Then for $P_+$ poles outside of the unit circle and $P_-$ poles inside the unit circle, $P_- + P_+ = P$, additive splits are defined as:
\begin{equation}
    f_-=\sum_{l=1}^{P_+}\frac{r_l^{+}}{z-z_p^{+l}}+r_0(z), \quad f_+ = \sum_{l=1}^{P_-}\frac{r_l^{-}}{z-z_p^{-l}}.
\end{equation}

The main difficulty of this step is to deal with non-proper rational functions and higher order poles, which arise when we are correspondingly multiplying or dividing a rational function by $z^p, \ p>1$. In the first case, we have to use a long division to obtain $r_0(z)$, whereas in the second case, we  need to solve a matrix equation to find residues associated to each $z^{-1},..,z^p$ term of the partial fraction decomposition. Each of these operations takes an inefficiently large amount of time for numerical calculations, so we are going to avoid them as follows. It is important to notice that multiple poles are always inside the unit circle and therefore should belong to the `plus' function, whereas the residual polynomial always belongs to the `minus' function since we want to restrict $f_+$ from algebraic growth at infinity. Thus, we have the following possible situations:
\begin{enumerate}
    \item If $f$ is a non-proper polynomial (multiplied by $z^p,\ p>0$, we calculate directly the positive component $f_+$ by \equ{eq:partfrac} and then find the minus component as:
    $$f_- = f - f_+.$$
    \item If $f$ possesses multiple poles (is divided by $z^p,\ p>0$) then, the other way around, we directly find $f_-$ from \equ{eq:partfrac} and then calculate $f_+$ as:
    $$f_+ = f - f_-.$$
    \item In case, when the function $f$ has no poles, we set $f_-=f,\ f_+ = 0$.
\end{enumerate}

Note that for each iteration, constant $C_2^{(\ell)}$  for \equ{eq:U_plus_even}-\equ{eq:U_minus_even} is determined at this step by \equ{eq:contant2}. 

\subsection{Results}
\label{sect:results_plots}
\begin{figure}[h]
    \centering
     \begin{subfigure}[b]{0.32\textwidth}
         \centering
         \caption{Re $u_{m,n}$}
         \includegraphics[width=\textwidth]{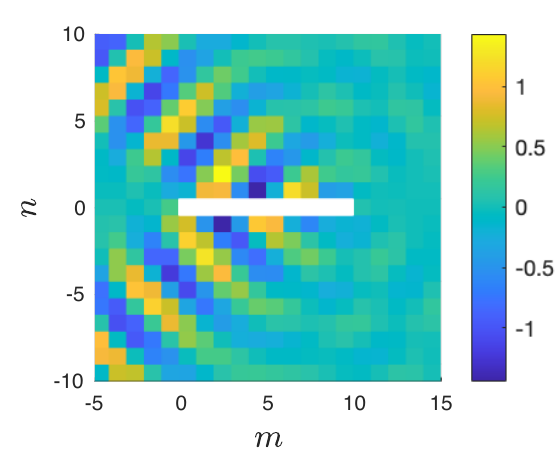} 
         \label{fig:heatmap_singcrack1}
     \end{subfigure}
    \begin{subfigure}[b]{0.32\textwidth}
        \centering
        \caption{Re $u_{m,n}^{tot}$}
        \includegraphics[width=\textwidth]{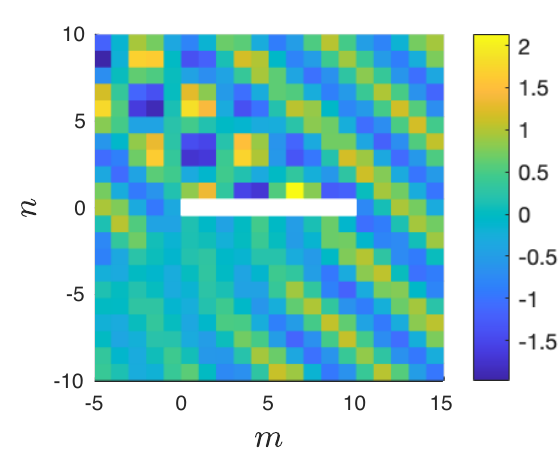} 
        \label{fig:heatmap_singcrack2}
    \end{subfigure}
     \begin{subfigure}[b]{0.32\textwidth}
         \centering
         \caption{$|u_{m,n}^{tot}|$}
         \includegraphics[width=\textwidth]{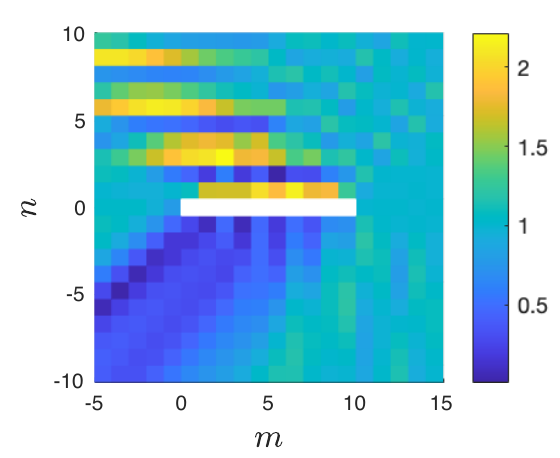} 
         \label{fig:heatmap_singcrack3}
     \end{subfigure}
    \centering
     \begin{subfigure}[b]{0.32\textwidth}
         \centering
         \caption{Re $u_{m,n}$}
         \includegraphics[width=\textwidth]{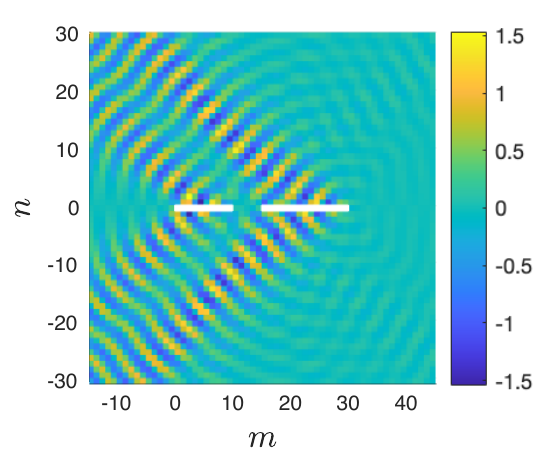} 
         \label{fig:heatmap_multcrack21}
     \end{subfigure}
     \begin{subfigure}[b]{0.32\textwidth}
         \centering
         \caption{Re $u_{m,n}^{tot}$}
         \includegraphics[width=\textwidth]{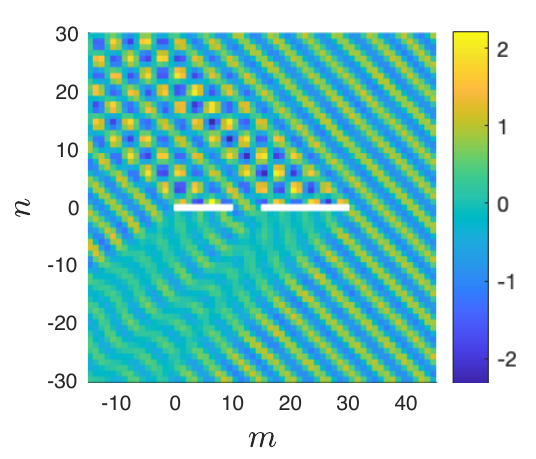} 
         \label{fig:heatmap_multcrack22}
     \end{subfigure}
     \begin{subfigure}[b]{0.32\textwidth}
         \centering
         \caption{$|u_{m,n}^{tot}|$}
         \includegraphics[width=\textwidth]{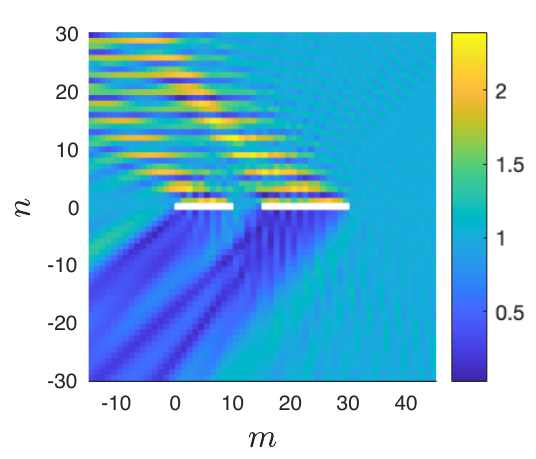} 
         \label{fig:heatmap_multcrack23}
     \end{subfigure}
    \centering
     \begin{subfigure}[b]{0.32\textwidth}
         \centering
         \caption{Re $u_{m,n}$}
         \includegraphics[width=\textwidth]{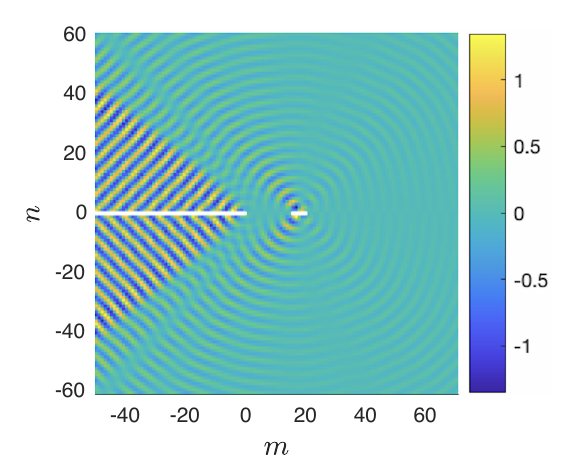} 
         \label{fig:heatmap_multcrack213}
     \end{subfigure}
     \begin{subfigure}[b]{0.32\textwidth}
         \centering
         \caption{Re $u_{m,n}^{tot}$}
         \includegraphics[width=\textwidth]{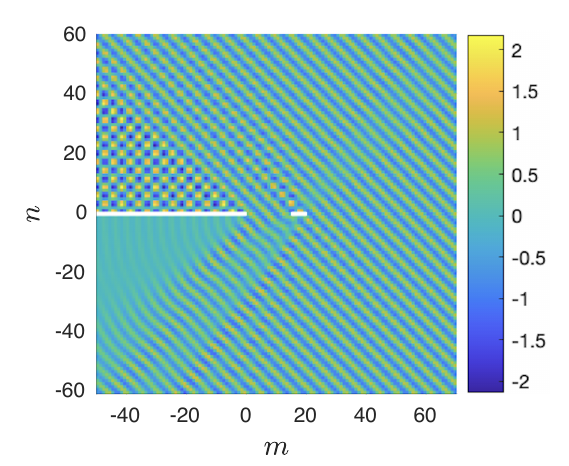} 
         \label{fig:heatmap_multcrack223}
     \end{subfigure}
     \begin{subfigure}[b]{0.32\textwidth}
         \centering
         \caption{$|u_{m,n}^{tot}|$}
         \includegraphics[width=\textwidth]{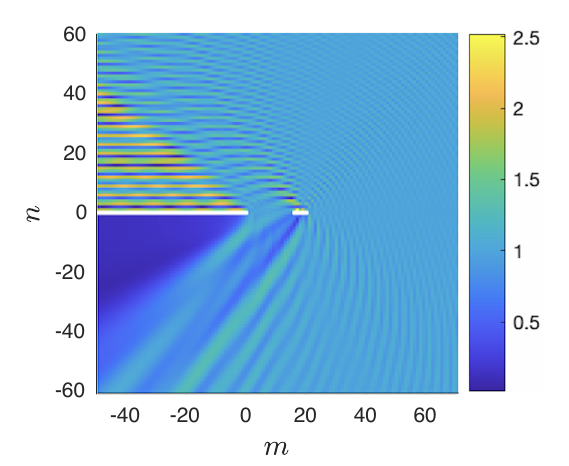} 
         \label{fig:heatmap_multcrack233}
     \end{subfigure}
\caption{
Displacement field on a square lattice for $K=0.5\pi,\ \Omega=1.49,\  \varphi^{\text{in}}=\pi/4$ in the case of a single crack of length 10 (a)-(c); 2 collinear cracks of lengths $10$ and $15$ (d)-(f); 2 collinear cracks, where one is semi-infinite and another is of length  $5$ (g)-(i).
}
\label{fig:heatmap_multcrack2}
\end{figure}

After the additive splitting is performed, the iterative procedure can be completed. By definition \equ{eq:transf_U1}, the full transform $U$ for the $j^{th}$ iteration is then restored from its components \equ{eq:minus}-\equ{eq:transf_even} for N collinear cracks as: 

\begin{equation}
    \left\{U\right\}^j = \left\{U_-^{(1)}\right\}^j +\sum_{\ell=1}^{2N}z^{-m_{\ell}}\left\{U^{(\ell)}_{+}\right\}^j.
    \label{eq:integrand_mult}
\end{equation}
Then, the scattered displacement field $\{u_{m,n}\}^j$ for the $j^{th}$ iteration can be found from the inverse Fourier transform \equ{eq:discr_inv_fourier}, and the total field is $\{u^{\text{tot}}_{m,n}\}^j = \{u_{m,n}\}^j + u_{m,n}^{\text{in}}$.
Note that the contour $\mathcal{C}_{z}$ for numerical integration of the discrete Fourier transform can be chosen to be the unit circle, when $\text{Im}\ \Omega \neq 0$, but for the limiting absorption case when $\text{Im}\ \Omega \to 0$ it has to be deformed to avoid the singularities of $\left\{U\right\}^j$ hitting the unit circle (\fig{fig:curved_contour}). In our case, the singularities are the poles in the rational approximation, that imitate the branch points of the original non-approximated function.
The resulting heatmaps for the $6^{th}$ iteration for a single crack are shown on \fig{fig:heatmap_multcrack2}. 

\FloatBarrier
\subsection{Convergence}
\begin{figure}[h]
\centering
    \begin{subfigure}[b]{0.45\textwidth}
         \centering
         \caption{Crack length $L=10$}
         \includegraphics[width=\textwidth]{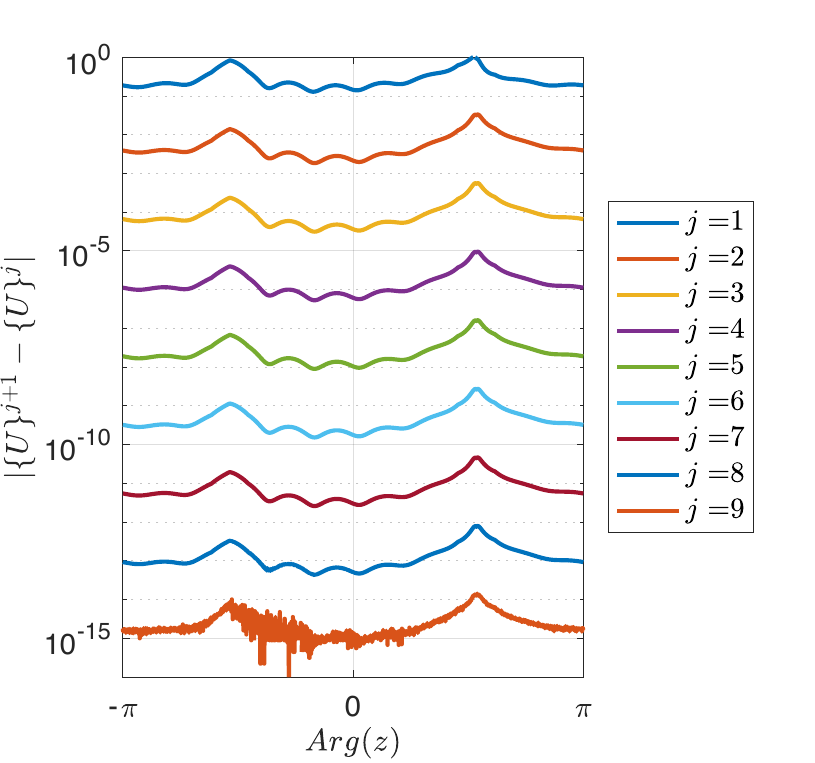}
         \label{fig:crack_conv1}
     \end{subfigure}
     \begin{subfigure}[b]{0.45\textwidth}
         \centering
         \caption{2 cracks of lengths $10$ and $15$}
         \includegraphics[width=\textwidth]{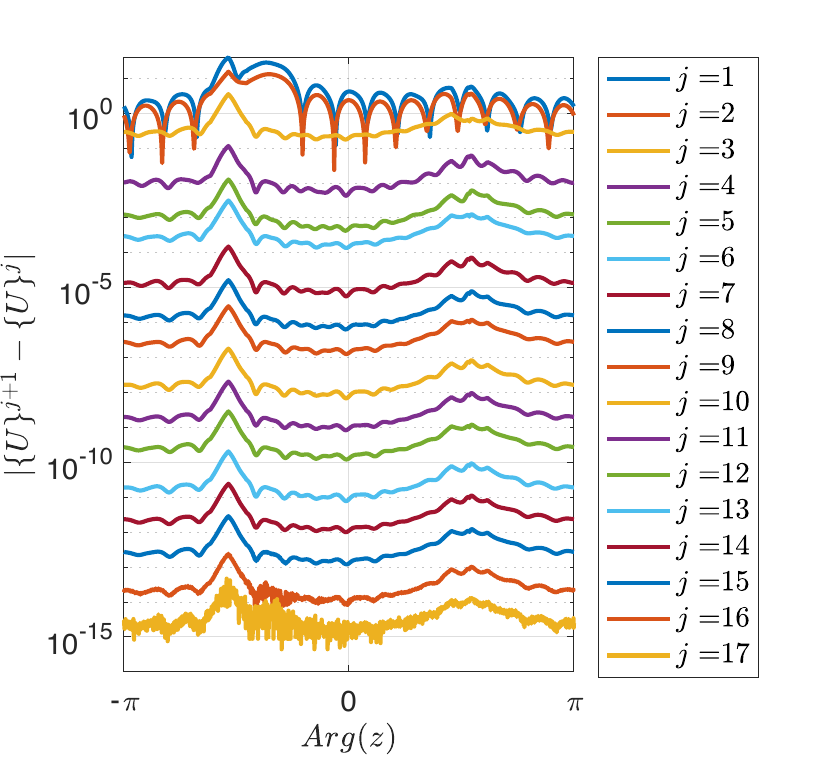}
         \label{fig:cracks_conv2}
     \end{subfigure}
    \caption{Absolute difference of spectral field between nearest iterations.}
        \label{fig:three_graphs}
\end{figure}

\begin{figure}[h]
\centering
    \begin{subfigure}[b]{0.49\textwidth}
        \centering
        \includegraphics[width=\textwidth]{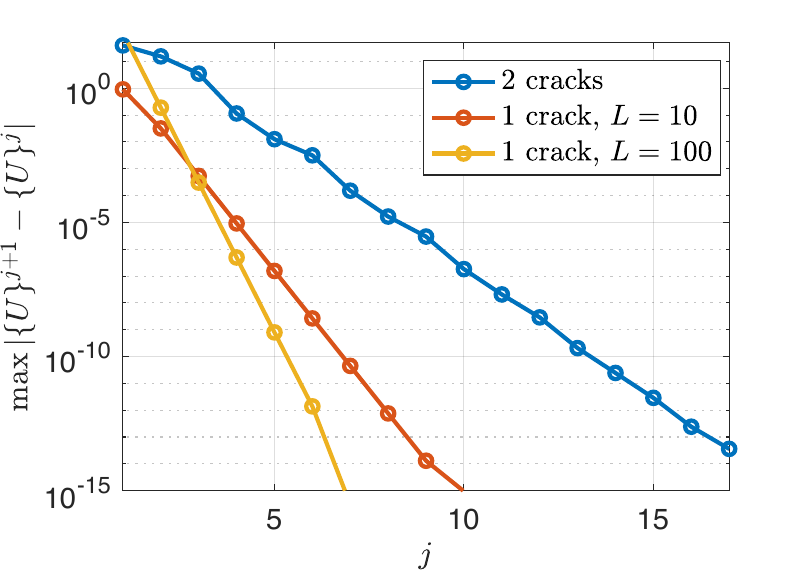} 
        \caption{}
        \label{fig:conv_compare}
    \end{subfigure}
    \begin{subfigure}[b]{0.48\textwidth}
        \centering
        \includegraphics[width=\textwidth]{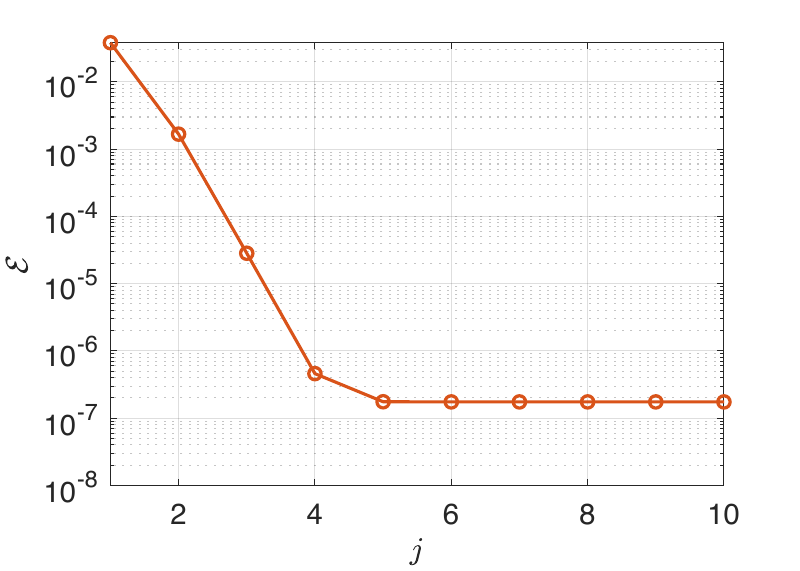} 
        \caption{}
        \end{subfigure}
    \caption{
    (a) Maximum absolute difference of spectral field between  nearest iterations; (b) Maximum absolute difference for the physical scattered field $\{u_{m,n}\}^j$ between the exact solution \equ{eq:ex_sol_crack} and the first 10 iterations for the single crack of the length 10.}
\end{figure}

It is shown in \fig{fig:three_graphs} and \fig{fig:conv_compare} that the iterations converge since their absolute difference on the unit circle decreases on each iterative step. 
One can see that a solution for a single crack converges within a few iterations to a machine precision, which is $\sim10^{-15}$ in \verb+Matlab+. This also happens for 2 cracks but takes more iterations.

The maximum absolute difference between iterations is shown in \fig{fig:conv_compare}. It shows that the long single crack demonstrates the fastest convergence whereas the algorithm converges significantly slower for two finite cracks. This result is predictable by the physics of multiple scattering on the crack edges: the longer the crack is the less interaction there is between its edges, and the more edges in the system there are to couple, the more iterations are required to achieve the desired accuracy. This is the opposite effect to the Green's function approach (see \sect{sect:Greens_function}) where the crack length significantly increases the computation time.
The maximum achievable accuracy can be estimated as the order of the rational approximation error, which is $10^{-7}$ in the considered case. Then, as we can conclude from \fig{fig:conv_compare}, five iterations are enough to achieve satisfactory accuracy for a single crack of length $L=10$. This observation is confirmed by comparing the iterative and exact solutions \equ{eq:ex_sol_crack}. There the error $\mathcal{E}(j)$ was estimated as:
$$\mathcal{E}(j)=\max\limits_{(m,n)\in\mathcal{D}} |\{u_{m,n}\}^j-u_{m,n}^{\text{exact}}|,$$
where $\mathcal{D}$ is a half-rectangular contour in the physical upper half-plane. Here $u_{m,n}^{\text{exact}}$ is defined as \equ{eq:ex_sol_crack}, where the discrete Green's functions are calculated as single integrals as defined in \cite{Shanin2020Sommerfeld-typeProblems} and described in \sect{sect:Greens_function}.

\FloatBarrier

%% file: 5_conclusions.tex
\section{Conclusions}
In this paper,  we have derived matrix Wiener-Hopf equations for the discrete problems of scattering by a single finite crack and a set of collinear cracks. By employing an iterative method alongside rational approximation, we avoid the need for matrix factorisation and reduce the problem to a sequence of scalar Wiener-Hopf equations. This represents an extension to the discrete setting of the iterative scheme designed for the continuous analogue of the problem
\cite{ Kisil2018AerodynamicExtensions, Priddin2020ApplyingFactors}. 

Our method was critically compared to other existing methods using Green's functions \cite{Sharma2015Near-tipCrack} for instance.
The advantages of the present approach include that its numerical complexity is virtually independent of the cracks' lengths, that the numerical complexity of an iteration does not depend on the iteration number, and that it can cater for a mixture of finite and semi-infinite cracks. Its limitation in terms of accuracy is related to the accuracy of the various rational approximations used, however, for a given rational approximation accuracy, the present method converges with a very modest number of iterations. In fact, fewer number of iterations are required for longer crack.


 The method has the potential to be extended to other types of lattices (e.g.\ triangular or honeycomb) and boundary conditions (e.g.\ rigid constraints as shown in \ref{appx:wh_rigid}), as well as to different kinds of Laplacian discretisation.

\section*{Acknowledgements}
 
A.V.K. is supported by a Royal Society Dorothy Hodgkin Research Fellowship and a Dame Kathleen Ollerenshaw Fellowship. E.M. is supported by a President’s Doctoral Scholarship Award of The University of Manchester. The authors would like to thank the Isaac Newton Institute for Mathematical Sciences for support and hospitality during the programme ‘Mathematical theory and applications of multiple wave scattering’ when work on this paper was undertaken. This programme was supported by EPSRC grant EP/R014604/1.

%% file: appendB.tex
\section{Region of analyticity for the discrete Fourier transform}
\label{app:appB}
To analyse the regions of convergence for $U_{n;+}$ and $U_{n;-}$ in a similar manner as in \cite{NobleB1958MethodsPDEs, Sharma2015DiffractionCrack}, we will note that the scattered field $u_{m,n}$ for $m\notin {M_n,L-M_n}$, where $M_n = -|n|\cot{\varphi^{\text{in}}}$, is a diffracted wave $u_{m,n}^{\text{d}}$, which is produced by the edge of the crack. Due to the 2D nature of the problem, $u_{m,n}^{\text{d}}$ is expected to have the form 
$$u_{m,n}^{\text{d}}\sim (m^2+n^2)^{-1/2}e^{iK(m^2+n^2)^{1/2}}.$$
Then from the fact that $\Omega$ and therefore $K$ has a small positive imaginary part $\varepsilon$ it follows:
\begin{align*}
    |u_{m,n}|<e^{-\varepsilon (m^2+n^2)^{1/2}} \quad \text{for} \quad m\in(-\infty,M_n)\cup(L-M_n,\infty).
\end{align*}
Hence for any $n$ we have $|u_{m,n}|<e^{-\varepsilon |m|}$ as $m\to\pm\infty$. Then $U_{n;+}$ is absolutely convergent and therefore analytic for $|z|>e^{-\varepsilon}$, $U_{n;-}$ for $|z|<e^{\varepsilon}$. 


%% file: appendA.tex
\section{Wiener-Hopf equation for finite rigid constraint in square lattice}
\label{appx:wh_rigid}

The rigid constraint is denoted as the set $\Sigma$ of the lattice sites that have fixed particles (\fig{fig:rigid_scheme}), such that:
\begin{equation}
    u_{m,0}^{\text{tot}}=0, \quad (m,n)\in\Sigma=\left\{(m, 0) \in \mathbb{Z}^{2}: 0<m<L \right\}.
    \label{eq:rigid_bc}
\end{equation}  
The problem of scattering of a plane incident field on a finite rigid constraint leads to a matrix Wiener-Hopf equation, similar to \equ{eq:matrixWH_crack}:
\begin{equation}
\left(\begin{array}{c} \left(U^{(1)}_{1}\right)_{-} \\ \\ \left(U_{1}^{(2)}\right)_-\end{array}\right) 
=
-\left(\begin{array}{cc}[\mathcal{K}^R(z)]^{-1} & z^{-L} \\ \\ -z^L & 0\end{array}\right) \left(\begin{array}{c}\left(U_{1}^{(2)}\right)_- \\ \\\left(U^{(2)}_{1}\right)_{+} \end{array}\right) 
+
\left(\begin{array}{c} F(z) \\ \\ 0\end{array}\right),
\label{eq:matrix_rigid}
\end{equation}
where $\mathcal{K}^R(z)$ and $F(z)$ are the known functions, given in \cite{Sharma2015DiffractionConstraint}. The derivation of \equ{eq:matrix_rigid} is similar to the procedure described in \cite{Sharma2015DiffractionConstraint} and the analogue for the finite crack (\sect{sect:WH_finite_crack}).